\documentclass[a4paper,11pt]{article}
\usepackage{jheppub} 
\usepackage{lineno}
\usepackage{graphicx}
\usepackage{amsmath,amsthm,amssymb,amsfonts,mathtools,bbold,hyperref}

\newcommand{\hs}{\mathcal{H}}

\arxivnumber{2304.12345} 

\title{\boldmath Non-isometric codes for the black hole interior from fundamental and effective dynamics}

\author{Oliver DeWolfe}
\author{and Kenneth Higginbotham}
\affiliation{Department of Physics and Center for Theory of Quantum Matter,\\
390 UCB University of Colorado Boulder, CO 80309, U.S.A.}

\emailAdd{oliver.dewolfe@colorado.edu, kenneth.higginbotham@colorado.edu}

\abstract{We introduce a new holographic map for encoding black hole interiors by including both fundamental and effective dynamics. This holographic map is constructed by evolving a state in the effective, semiclassical gravity description of the interior backwards in time to pull the degrees of freedom outside the black hole, before evolving forwards in time in the fundamental description. We show this ``backwards-forwards'' map is equivalent to a post-selection map of the type introduced by Akers, Engelhardt, Harlow, Penington, and Vardhan, and in the case of trivial effective interactions reduces to their model, while providing a suitable generalization when those interactions are nontrivial. We show the map is equivariant with respect to time evolution, and independent of any interactions outside the black hole. This construction includes interactions with an infaller in a way that preserves the unitarity of  black hole evolution exactly and does not allow for superpolynomial computational complexity.}

\begin{document}
\maketitle
\flushbottom

\section{Introduction}

The study of quantum gravity through problems such as the black hole information paradox has received a great deal of recent interest, and for good reason: the consequences of Hawking's discovery that black holes evaporate \cite{hawking_particle_1975} provided an excellent test bed for pushing the limits of general relativity and quantum mechanics. Here the AdS/CFT correspondence \cite{Maldacena:1997re,Gubser:1998bc,Witten:1998qj} has proved particularly useful, providing a holographic map that directly relates gravitational phenomena in the bulk of AdS to field theory processes on the CFT boundary. Generally this map embeds a subspace of the bulk Hilbert space in the boundary isometrically. This has led to many connections with quantum error correction which have proven fruitful in pursuits of a resolution to the paradox (see for example \cite{almheiri_bulk_2015,dong_reconstruction_2016,harlow_ryu-takayanagi_2017}).

The interiors of black holes appear to pose an issue to this picture. From the outside of a black hole, the area of the event horizon counts the fundamental quantum gravity degrees of freedom contained within; the exact nature of this \textit{fundamental description} of the black hole states is unknown, but in the case of AdS/CFT they may be identified with boundary degrees of freedom. As the black hole forms and then evaporates, the number of degrees of freedom in the fundamental description increases and then decreases to zero. An infalling observer, however, sees a different picture, continuing to describe the local physics, at least approximately, with an \textit{effective description} of the black hole given by bulk semiclassical gravity. The interior continuously grows with time, even as the horizon shrinks, and thus the number of effective field theory modes inside the black hole eventually dwarfs the number of actual black hole degrees of freedom. The holographic map relating the effective description of the black hole interior to the fundamental description needs to annihilate some states to reduce the size of the Hilbert space, making it non-isometric.

Recent work by Akers, Engelhardt, Harlow, Penington, and Vardhan (PHEVA\footnote{We found this permutation of the authors' initials particularly user-friendly.}) \cite{akers_black_2022} has attempted to characterize such a non-isometric holographic map. The key characteristic of their construction is the use of \textit{post-selection} as an essential part of the map. This model  had great success in reproducing key results of the field, including a derivation of the QES formula \cite{ryu_holographic_2006,faulkner_quantum_2013,engelhardt_quantum_2015,penington_entanglement_2020,almheiri_entropy_2019} that has led to an incredible amount of recent progress in understanding the information paradox, and a demonstration that the non-isometric aspects of the map and the breakdown of the effective description are only visible to processes of exponential computational complexity \cite{harlow_quantum_2013,brown_pythons_2019}. These and other successes have inspired a great deal of interest in non-isometric holographic maps encoding black hole interiors. Many avenues of research following the work of \cite{akers_black_2022} are already under way; see for example \cite{antonini_holographic_2022,
kar_non-isometric_2022,
antonini_holographic_2023,
faulkner_asymptotically_2022,
giddings_comparing_2022,
gyongyosi_holographic_2023,
chandra_toward_2023,
yang_complexity_2023,
cao_approximate_2023,
kubicki_constraints_2023}.

The authors of \cite{akers_black_2022} also discuss how a holographic map involving post-selection can arise from the fundamental dynamics of the black hole. They demonstrate their construction is equivariant, in the sense that effective description time evolution followed by the holographic map is identical to the holographic map followed by fundamental description time evolution. In constructing their model, PHEVA assumes that the black hole dynamics in the effective description are trivial. There are no interactions between degrees of freedom inside the black hole, nor do infalling degrees of freedom interact with anything on their journey past the horizon. Recent work by Kim and Preskill \cite{kim_complementarity_2022} has begun to generalize this by considering how an infaller can interact with modes inside and outside of the black hole. Motivated by questions surrounding the final state proposal \cite{horowitz_black_2004,lloyd_unitarity_2014},  this work verified that these new interactions do not combine with the post-selection in the holographic map proposed by PHEVA to give large deviations from unitarity in the black hole's evolution. Furthermore, they showed that the post-selection does not lead to any superpolynomial computational complexity that would violate the quantum extended Church-Turing thesis \cite{aaronson_quantum_2005,deutsch_quantum_1997,susskind_horizons_2020} or other problems typically associated with post-selection.

However, there is a subtlety in the post-selection holographic map that requires greater care when adding additional interactions in the effective description. Once more general effective dynamics and interactions are included, the effective degrees of freedom cannot simply be fed into the holographic map as it is currently constructed; the map requires degrees of freedom in the fundamental description. Motivated by the need to include more general effective dynamics, we propose a new formulation of the holographic map encoding black hole interiors. Our construction implements the holographic map as a composition of effective dynamics backwards in time, followed by fundamental dynamics forwards in time. This ``backwards-forwards'' holographic map takes the effective description at a moment in time, runs its evolution backwards to before the black hole was formed, and then runs the evolution forward in the fundamental description. This approach is manifestly equivariant, and it can incorporate interactions, both of semiclassical modes inside the horizon, and of interactions between the infalling modes and outgoing Hawking radiation outside the black hole. Moreover, we show that this seemingly orthogonal construction can be easily and naturally recast as a post-selection holographic map as proposed by \cite{akers_black_2022},
at least for states formed by unitary evolution of smooth matter falling into the black hole.
When the interactions are removed, it reduces to the map of \cite{akers_black_2022}; when interactions are included, it represents an appropriate generalization. 

We are also able to show that this holographic map acts trivially on the reservoir outside the black hole even in the presence of reservoir interactions, for example between infalling modes and outgoing Hawking radiation; in fact the map is independent of any such interactions. We argue that the map is exactly unitary when acting on a valid black hole state. Moreover, since the post-selection form of the map can be related to the backwards-forwards form that lacks nontrivial post-selection, no superpolynomial computational complexity should arise.

The remainder of this paper will be structured as follows. Section~\ref{sec:review} will present a brief review of the non-isometric post-selection holographic map of \cite{akers_black_2022}, and its relation to a simple model of black hole dynamics. In section~\ref{sec:BFmap}, we generalize the black hole dynamics by adding interactions, and introduce the manifestly equivariant backwards-forwards holographic map involving a composition of the dynamics in the effective and in the fundamental description. We show how to recast this map in a post-selection form, and show that it reduces to the case studied by PHEVA in the proper limit. In section~\ref{sec:outside}, we add interactions outside of the black hole and show that they drop out of the holographic map, and that equivariance continues to hold. We also consider the questions posed by Kim and Preskill in the context of our new holographic maps, showing that the backwards-forwards map provides a simple resolution to those issues. Finally, we make some concluding remarks in section~\ref{sec:conc}.

\section{Review of post-selection holographic map} \label{sec:review}

We begin with a brief review of the non-isometric holographic map proposed by PHEVA \cite{akers_black_2022}. Here, we will focus on the structure of the holographic map and its relation to fundamental dynamics. Other important properties and applications of the code (such as the derivation of the QES formula and the role of complexity) will not play a role in this work, so we refer the interested reader to \cite{akers_black_2022} for more details.

\subsection{Two descriptions of a black hole and the holographic map}

The holographic map relates two different presentations of the black hole and its dynamics, the \textit{effective description} and the \textit{fundamental description}. In both cases, the black hole is surrounded by a ``reservoir" $R$, from which modes fall into the horizon, and into which the black hole radiates.

The \textit{effective description} models a black hole in the semiclassical approximation. The geometry of spacetime is fully present, and there are effective field theory modes living on and influencing this geometry. The Hilbert space may be factorized as
\begin{equation}
    \hs = \hs_\ell \otimes \hs_r \otimes \hs_{R_\text{in}} \otimes \hs_{R_\text{out}}\,.
\end{equation}
The $\ell$ and $r$ modes exist inside the black hole, describing left-moving (radially ingoing) and right-moving (radially outgoing) modes, respectively. Degrees of freedom outside the black hole are contained in $R_\text{in}$, describing any infalling modes while they are still outside the horizon, and $R_\text{out}$, containing the Hawking radiation created during the evaporation of the black hole. Hawking modes in $R_\text{out}$ are created entangled with corresponding modes in $r$. This effective description of the black hole is valid at low curvature and complexity, and we consider it to be the account given by an infalling observer in these regimes of validity.

An outside observer provides the second description of the black hole, called the \textit{fundamental description}. Here, fundamental quantum gravity degrees of freedom $B$ (presently unknown to us in a bulk description, and perhaps best modeled by dual CFT degrees of freedom) are used to describe the black hole. The same reservoir $R$ provides two more tensor factors for the degrees of freedom outside the black hole and completes the Hilbert space:
\begin{equation}
    \hs = \hs_B \otimes \hs_{R_\text{in}} \otimes \hs_{R_\text{out}}.
\end{equation}
The geometry of the black hole is not present in this description, as we take geometry to be an emergent property in quantum gravity. 

The holographic map $V$ is a map from the effective degrees of freedom $\ell$ and $r$ inside the black hole to the fundamental $B$ degrees of freedom,
\begin{equation}
    V \,:\, \hs_\ell \otimes \hs_r \rightarrow \hs_B.
\end{equation}
Because the same reservoir is shared by both descriptions, the holographic map acts trivially on $\hs_R$. The map described by PHEVA begins by tensoring in degrees of freedom $f$ in a fixed state that keeps track of any modes red-shifted down from above the cutoff scale during the collapse to form the black hole or other degrees of freedom held fixed. A unitary transformation $U$ is then performed on $\ell rf$, rearranging the degrees of freedom into $B$ and a temporary tensor factor called $P$. The excess of effective field theory modes in $\ell r$ at late times requires $V$ to be non-isometric, reducing the size of the Hilbert space to match $|B|$. To accomplish this, the authors of \cite{akers_black_2022} proposed that $V$ finishes by post-selecting on the subsystem $P$. Altogether, $V$ may be expressed as
\begin{equation}
    V = \sqrt{|P|} \, \langle \phi|_P U |\psi\rangle_f,
\end{equation}
where $|\psi\rangle_f$ indicates the insertion of the fixed state, and $\langle\phi|_P$ represents post-selecting $P$ on some specified state, with the numerical factor $\sqrt{|P|}$ fixing the normalization after post-selection.

\begin{figure}
    \centering
    \includegraphics[width=0.4\linewidth]{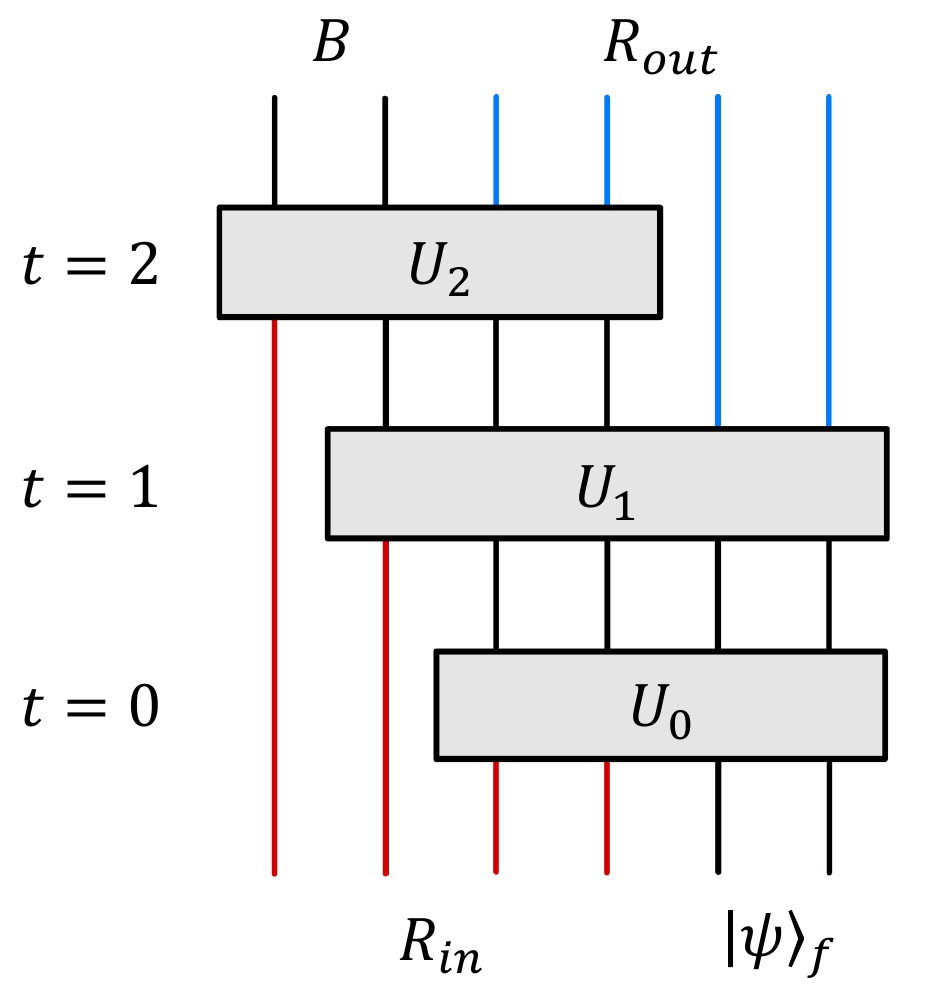}
    \caption{An example of fundamental dynamics in the PHEVA model for $n_0=4$ and $m_0=2$ up until $t=2$. Lines are colored by part of the Hilbert space the qudit lives in: black for $f$ and $B$, red for $R_\text{in}$, and blue for $R_\text{out}$; we retain this color scheme in later figures. Black boxes denote fundamental unitary dynamics $U_t$.}
    \label{fig:fun_dyn_PHEVA}
\end{figure}

\subsection{Relationship to dynamics}
The work of \cite{akers_black_2022} also constructs an implementation of this post-selection map using the dynamics of the black hole. They model all the degrees of freedom in terms of a set of qudits, and do not attempt to implement geometrical features like local Lorentz invariance. The dynamics in the fundamental description are modeled by a sequence of unitaries $U_t$ acting at each time step $t$. The black hole is formed at $t=0$ from $m_0$ qudits falling in from $R_\text{in}$ and $n_0 - m_0$ qudits in $f$; the unitary $U_0$ converts these to $n_0$ qudits in $B$, 
\begin{equation}
    U_0\,:\,R_\text{in}^{(m_0)} f^{(n_0-m_0)} \rightarrow B^{(n_0)},
\end{equation}
where parenthetical superscripts have been used to denote the number of qudits in each factor.\footnote{As is done in \cite{akers_black_2022}, we will take the black hole to be formed by a ``fast collapse,'' requiring $n_0 \gg m_0$. In the figures that follow we will not show this assumption directly, instead reducing the number of $f$ lines to make our diagrams more readable.} At each subsequent time step, the model assumes that one degree of freedom falls past the horizon, passing from $R_\text{in}$ to $B$, and two are released by the black hole as Hawking radiation, passing from $B$ to $R_\text{out}$. This is modeled by the unitary $U_t$ acting on all qudits in $B$ and one additional qudit from $R_\text{in}$, releasing as output two qudits into $R_\text{out}$ while the rest remain in $B$. $U_t$ thus acts as
\begin{equation}
    U_t\,:\, R_\text{in}^{(1)} B^{(n_0 + 1 - t)} \rightarrow B^{(n_0 - t)} R_\text{out}^{(2)}.
\end{equation}
The black hole shrinks by one qudit after every time step, completely evaporating after $n_0$ time steps. Figure~\ref{fig:fun_dyn_PHEVA} shows an example of these dynamics for $n_0 = 4$ and $m_0 = 2$ until $t=2$, where colors of the lines corresponding to qudits indicate which factor of the Hilbert space they live in:
black for $f$ and $B$, red for $R_\text{in}$, and blue for $R_\text{out}$. We retain these color conventions throughout the paper.

\begin{figure}
    \centering
    \includegraphics[width=0.7\linewidth]{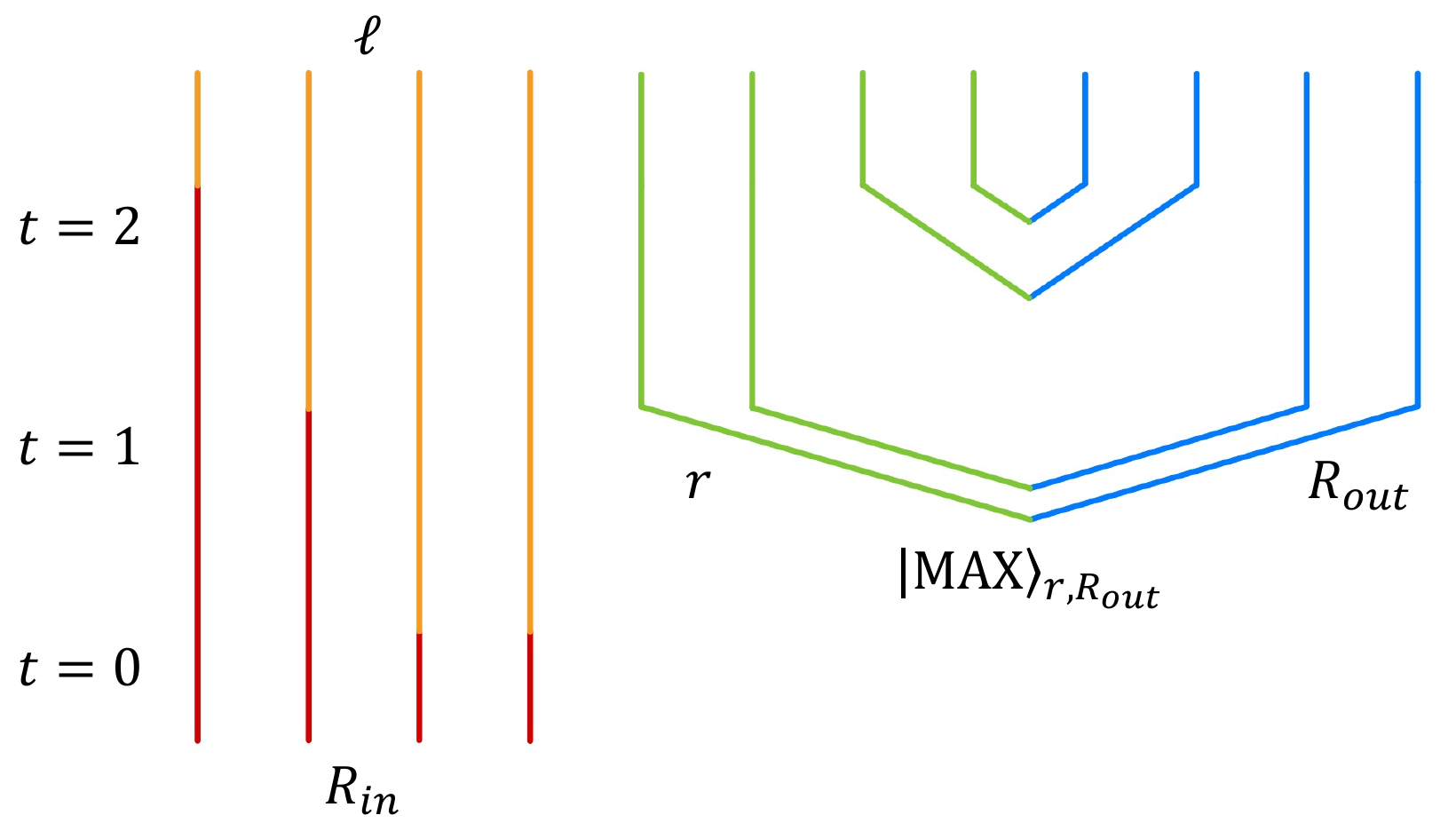}
    \caption{The effective dynamics in the PHEVA model up until $t=2$ for the same situation as figure~\ref{fig:fun_dyn_PHEVA}. Lines are colored by qudit type: orange for $\ell$, green for $r$, red for $R_\text{in}$, and blue for $R_\text{out}$; we retain this color scheme in later figures. $|\text{MAX}\rangle_{r,R_\text{out}}$ denotes the insertion of a maximally entangled pair of qudits on $rR_\text{out}$.}
    \label{fig:eff_dyn_PHEVA}
\end{figure}

The dynamics in the effective description must necessarily take in the same number of qudits from $R_\text{in}$ and emit the same number of qudits to $R_\text{out}$ at each time step as the fundamental dynamics, since the action on the reservoir $R$ is the same. At $t=0$, $m_0$ qudits pass from $R_\text{in}$ into $\ell$, constituting the formation of the black hole. At each subsequent time step, one more qudit falls into the black hole and is taken from $R_\text{in}$ and added to $\ell$. At the same time, two sets of maximally entangled Hawking pairs are created. One qudit from each pair exists behind the horizon and joins $r$, while the other escapes as Hawking radiation and is added to $R_\text{out}$. Outside of adding new qudits to each tensor factor, PHEVA takes the effective dynamics of $\ell rR_\text{out}$ to be trivial; no unitaries are applied at each time step. Figure~\ref{fig:eff_dyn_PHEVA} shows an example of these effective dynamics for two time steps, beginning with $m_0 = 2$. Again we distinguish the factors of the Hilbert space the qudits live in by the color of the lines: orange for $\ell$ and green for $r$, as well as red for $R_\text{in}$ and blue for $R_\text{out}$ as before, retaining these color conventions throughout the paper.

\begin{figure}
    \centering
    \includegraphics[width=0.7\linewidth]{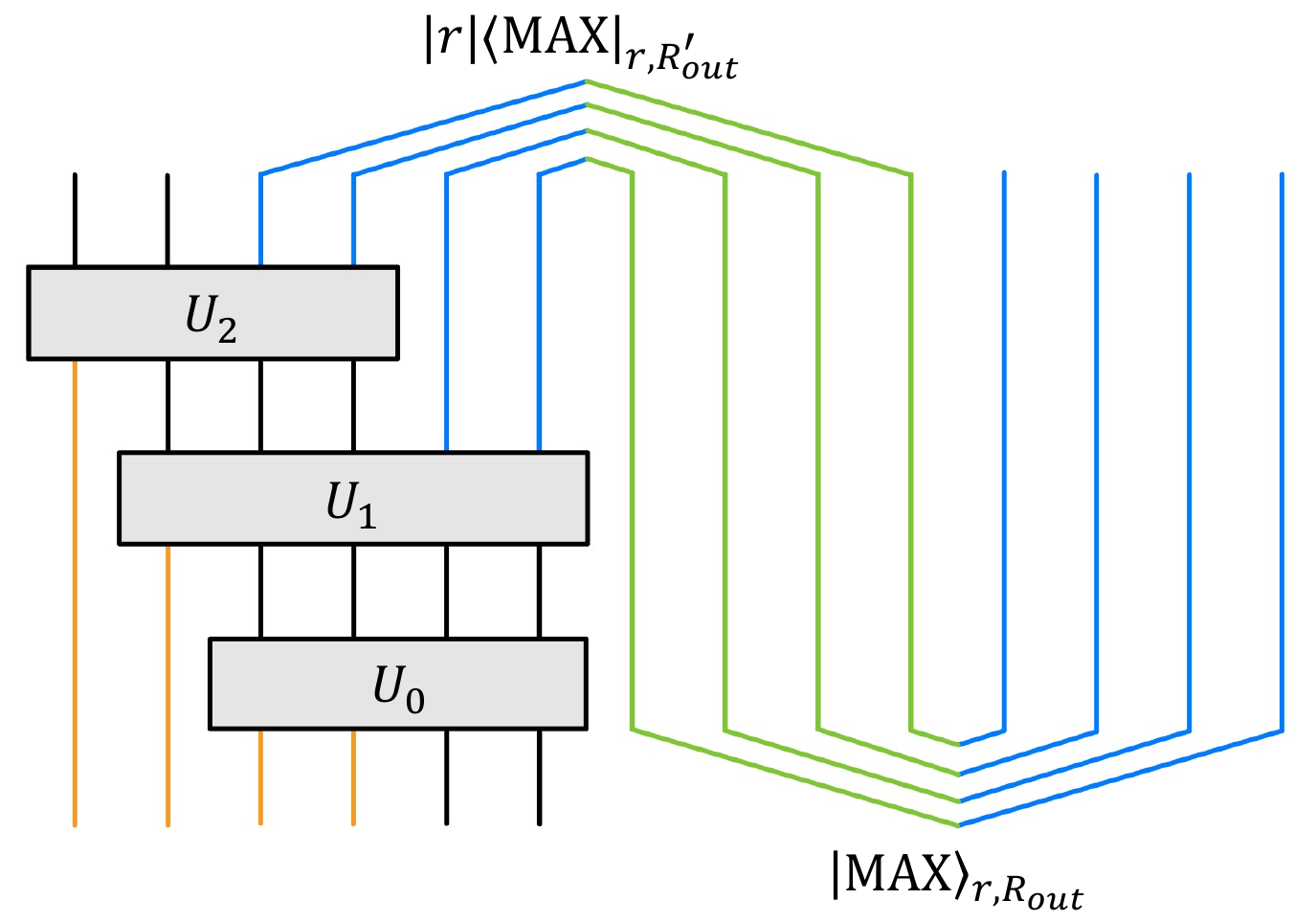}
    \caption{The non-isometric holographic map of PHEVA at $t=2$ relating the effective description shown in figure~\ref{fig:eff_dyn_PHEVA} to the fundamental description shown in figure~\ref{fig:fun_dyn_PHEVA}, including the teleportation protocol and post-selection on $rR'_\text{out}$.}
    \label{fig:V_PHEVA}
\end{figure}

At any given time $t$, one may consider the holographic map $V_t$ that takes the state of the black hole in the effective description and turns it into a state of the black hole in the fundamental description at the same time. PHEVA gives an explicit construction of their non-isometric holographic map  using the fundamental dynamics described above. Acting on the states of $\ell$, $r$, and $R_\text{out}$ in the effective description at time $t$, $V_t$ first appends the fixed degrees of freedom $f$. It then applies a unitary transformation $U$ to the subsystem $\ell f$. Here $U$ is defined by the fundamental time dynamics of the black hole,
\begin{equation}
    U = U_t U_{t-1}\dots U_0,
\end{equation}
converting the effective degrees of freedom to states on $B$ and a second copy of $R_\text{out}$, which we call $R'_\text{out}$. (We may identify these $R'_\text{out}$ modes with the temporary Hilbert space $P$ mentioned above.) The original $r$ and $R_\text{out}$ modes from the effective description are still present. Finally, post-selection is performed on $R'_\text{out}$  being in the maximally entangled state with the Hawking partner qudits in $r$, leaving a state involving degrees of freedom in $B$ and $R_\text{out}$ only and representing the state of the system in the fundamental description. This holographic map may be written as
\begin{equation}
    V_t =  |r| \langle \text{MAX} |_{r, R'_\text{out}} U |\psi\rangle_f \,,
\end{equation}
 with $\langle \text{MAX}|_{r,R_\text{out}}$ denoting the post-selection of $rR_\text{out}$ on the maximally entangled state, and the factor of $|r|$ fixing the normalization after this post-selection. Figure~\ref{fig:V_PHEVA} shows an example of this non-isometric map.

 One may view the ``bent'' lines connecting the output of $U$ to $R_\text{out}$ as forming a quantum teleportation protocol, teleporting the information inside the black hole to the radiation:
 \begin{equation}
 \label{eq:Teleportation}
     |r| \langle \text{MAX} |_{r, R'_\text{out}} \left( |\Psi\rangle_{R'_\text{out}} \otimes |\text{MAX}\rangle_{r, R_\text{out}} \right)  = |\Psi\rangle_{R_\text{out}} \,,
 \end{equation}
 for any state $|\Psi \rangle$ coming out of the unitaries $U_t$.
 Thus the ``bent" lines can be straightened to reveal the fundamental dynamics of the black hole.

 The authors of \cite{akers_black_2022} also demonstrate that this holographic map is equivariant, in the sense that waiting until time step $t+1$ (still assuming trivial effective dynamics) and then acting with $V_{t+1}$ is equivalent to acting with $V_t$ and then evolving with the fundamental dynamics $U_{t+1}$:
\begin{equation}
    V_{t+1} = U_{t+1} V_t\,.
\end{equation}
This relation follows immediately from ``straightening" the bent lines to remove the teleportation protocol, rewriting the map in terms of the fundamental dynamics, appending the extra factor $U_{t+1}$ to the dynamics, and then restoring the bent lines of the teleportation protocol at the new time step.

Let us now note a subtlety arising from this definition of the holographic map, which was part of the motivation for this work. The unitaries $U_t$ describing the fundamental dynamics take as inputs infalling modes in $R_\text{in}$ and fundamental degrees of freedom living in $B$, not the effective degrees of freedom $\ell$. How then can we feed modes in $\ell$ into a holographic map whose unitary is constructed from fundamental dynamics, as in figure~\ref{fig:V_PHEVA}? For the model considered in \cite{akers_black_2022}, the answer lies in the assumed triviality of the effective dynamics: since no unitaries were applied to $\ell$ in the effective description, there is a trivial correspondence between $\ell$ and $R_\text{in}$ modes. We may easily imagine pulling the $\ell$ degrees of freedom out of the black hole, returning them to $R_\text{in}$ without affecting the state. Since $U_t$ does take $R_\text{in}$ modes as an input, we may feed them directly into the holographic map without issue.\footnote{We thank Chris Akers and Daniel Harlow for discussions on this point.}

However, if we wish to include additional effective dynamics for $\ell rR$, we must deal with this subtlety more carefully to define the holographic map. There will no longer be a trivial correspondence between modes in $\ell$ and the $R_\text{in}$ modes which fell in to the black hole to create them, since they will have experienced interactions in the meantime. Entanglement originally between $r$ and $R_\text{out}$ may have become shared with $\ell$. We must do more to bring $\ell$ back to $R_\text{in}$ before performing the fundamental dynamics. The remainder of this work will be dedicated to developing a new construction for these holographic maps that can include nontrivial effective dynamics. As we shall see, the ``backwards-forwards" map we describe can be recast as a ``post-selection" map of the kind developed by \cite{akers_black_2022}, while being able to take interactions into account.

\section{The ``backwards-forwards'' holographic map} \label{sec:BFmap}

In the previous section, we found that using fundamental dynamics alone to construct the holographic map is not enough if we wish to include additional interactions between the $\ell$, $r$, and $R$ modes of the effective description. Here, we develop an approach to constructing a holographic map including these new effective dynamics. Throughout this section we will continue to use the fundamental dynamics described in section~\ref{sec:review} and will refer to figure~\ref{fig:fun_dyn_PHEVA} as our example for them. Here we focus on effective dynamics behind the horizon; the possibility of interactions outside the black hole are studied in the next section.

\subsection{Nontrivial effective dynamics behind the horizon}

We begin by defining nontrivial dynamics for $\ell$ and $r$ in the effective description, allowing for general unitary interactions $\hat{U}_t$ at each time step as degrees of freedom fall into the black hole. At $t=0$, $m_0$ qudits from $R_\text{in}$ cross the horizon; we allow for the possibility of interactions between the modes at this step by acting on the modes by a unitary transformation $\hat{U}_0$ as they pass into $\ell$,
\begin{equation}
    \hat{U}_0 \,:\, R_\text{in}^{(m_0)} \rightarrow \ell^{(m_0)}.
\end{equation}
At each following time step, one qudit from $R_\text{in}$ crosses the event horizon, and we allow interactions between it and any \textit{pre-existing} modes in $\ell$ and $r$, as well as interactions of the pre-existing $\ell$ and $r$ modes amongst themselves. 
We model all of these new interactions with the unitary $\hat{U}_t$ applied to $\ell r$ and one qudit from $R_\text{in}$,
\begin{equation}
    \hat{U}_t \,:\, R_\text{in}^{(1)} \ell^{(m_0 - 1 + t)} r^{(2(t-1))} \rightarrow \ell^{(m_0 + t)} r^{(2(t-1))}.
\end{equation}
In addition, we again create two sets of maximally entangled Hawking pairs in $rR_\text{out}$; the newly created modes in $r$ can interact with other modes beyond the horizon at future time steps.

\begin{figure}
    \centering
    \includegraphics[width=0.6\linewidth]{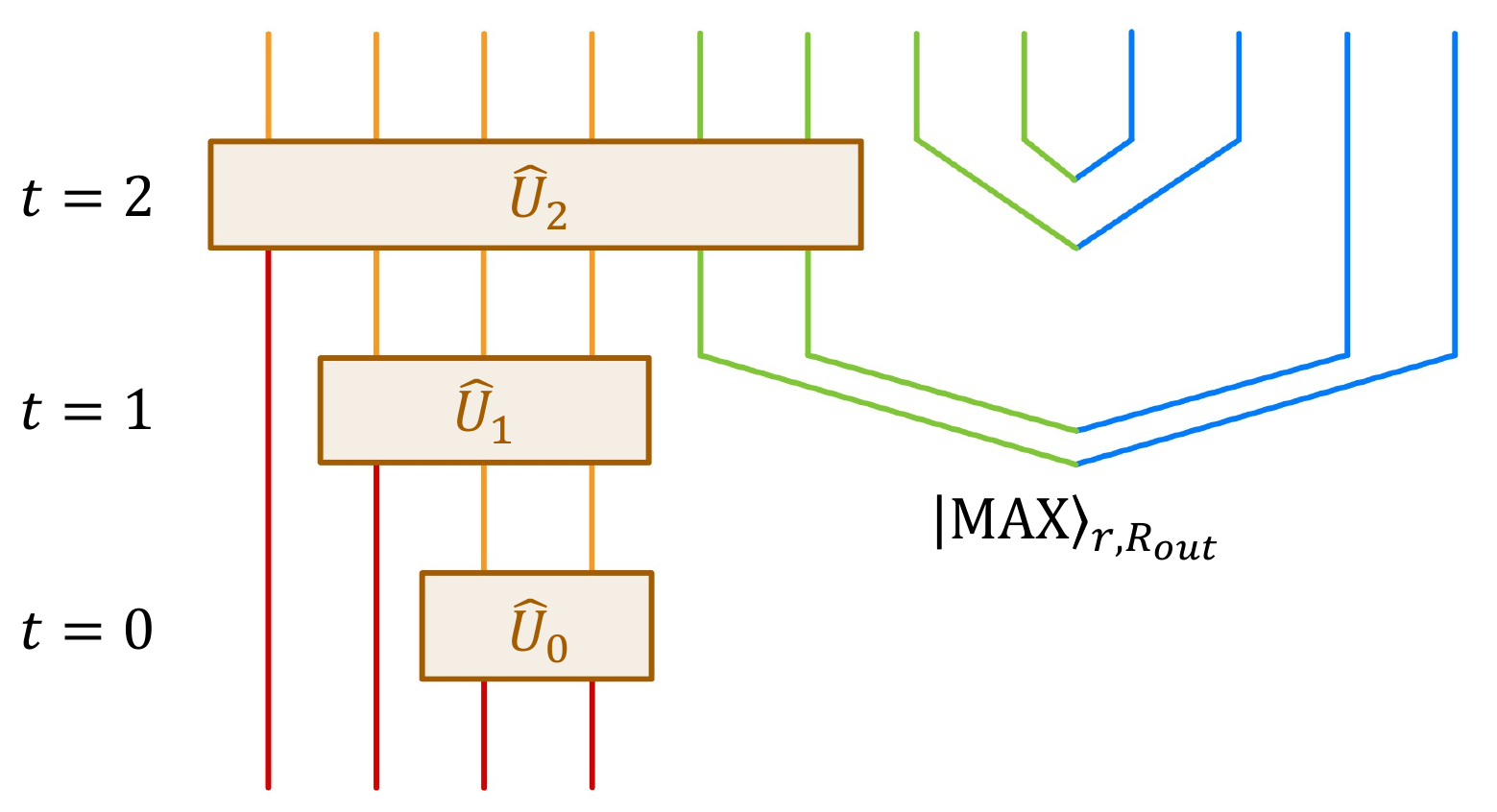}
    \caption{Nontrivial effective dynamics added to the example shown in figure~\ref{fig:eff_dyn_PHEVA}. Brown boxes denote effective unitary dynamics $\hat{U}_t$.}
    \label{fig:eff_dyn_NEW}
\end{figure}

Figure~\ref{fig:eff_dyn_NEW} shows an example of these dynamics for $m_0=2$ up until $t=2$. We note that after this time step, the entanglement that $R_\text{out}$ originally shared with $r$ has spread into $\ell$ as well; the Hawking radiation is still entangled with modes beyond the horizon, but interactions can spread this entanglement around.

We note that in general, we would expect these effective dynamics to have properties reflecting the semiclassical gravity description like (at least approximate) locality, so that some modes would interact more significantly with nearby modes and so on. 
We will not need to specify these properties of $\hat{U}_t$ in this work, so we leave them as general unitary transformations.

\subsection{Building the backwards-forwards holographic map} \label{sec:build}
Now that we have described new dynamics to capture interactions among the effective degrees of freedom, we construct a holographic map that is compatible with both these and the fundamental dynamics of section~\ref{sec:review}. The construction we describe here relies on the two descriptions sharing the same reservoir outside the black hole; if we can bring all the degrees of freedom to the reservoir, the two can be matched. Bringing the effective degrees of freedom inside the black hole back out involves evolution backwards in time.  Thus our implementation of the holographic map involves time evolution backwards in the effective description, followed by time evolution forwards in the fundamental description.

We begin with the state of $\ell rR_\text{out}$ at time $t$. Time evolving backwards in the effective description, we reverse all of the dynamics and interactions described above by actions of $\hat{U}_t^\dagger$, $\hat{U}_{t-1}^\dagger$, $\ldots \hat{U}_0^\dagger$.\footnote{One must be careful to undo all of the interactions perfectly to avoid creating any past singularities \cite{akers_black_2022}.} At each reversed time step, one qudit in $\ell$ emerges from behind the horizon and is returned to $R_\text{in}$. Additionally, at each step entanglement of $R_\text{out}$ that spread into $\ell$ modes is returned to the Hawking partners in $r$, leaving pairs of $rR_\text{out}$ modes in the maximally entangled state; we may think of these Hawking modes as annihilating and returning to the vacuum, and we can remove them by acting with the state $\langle \text{MAX}|_{r, R_\text{out}}$, which acts with unit overlap since all Hawking modes have returned to this state by definition. After all effective dynamics have been reversed, the radiation and black hole are gone, leaving all qudits in $R_\text{in}$.

\begin{figure}[!]
    \centering
    \includegraphics[width=0.7\linewidth]{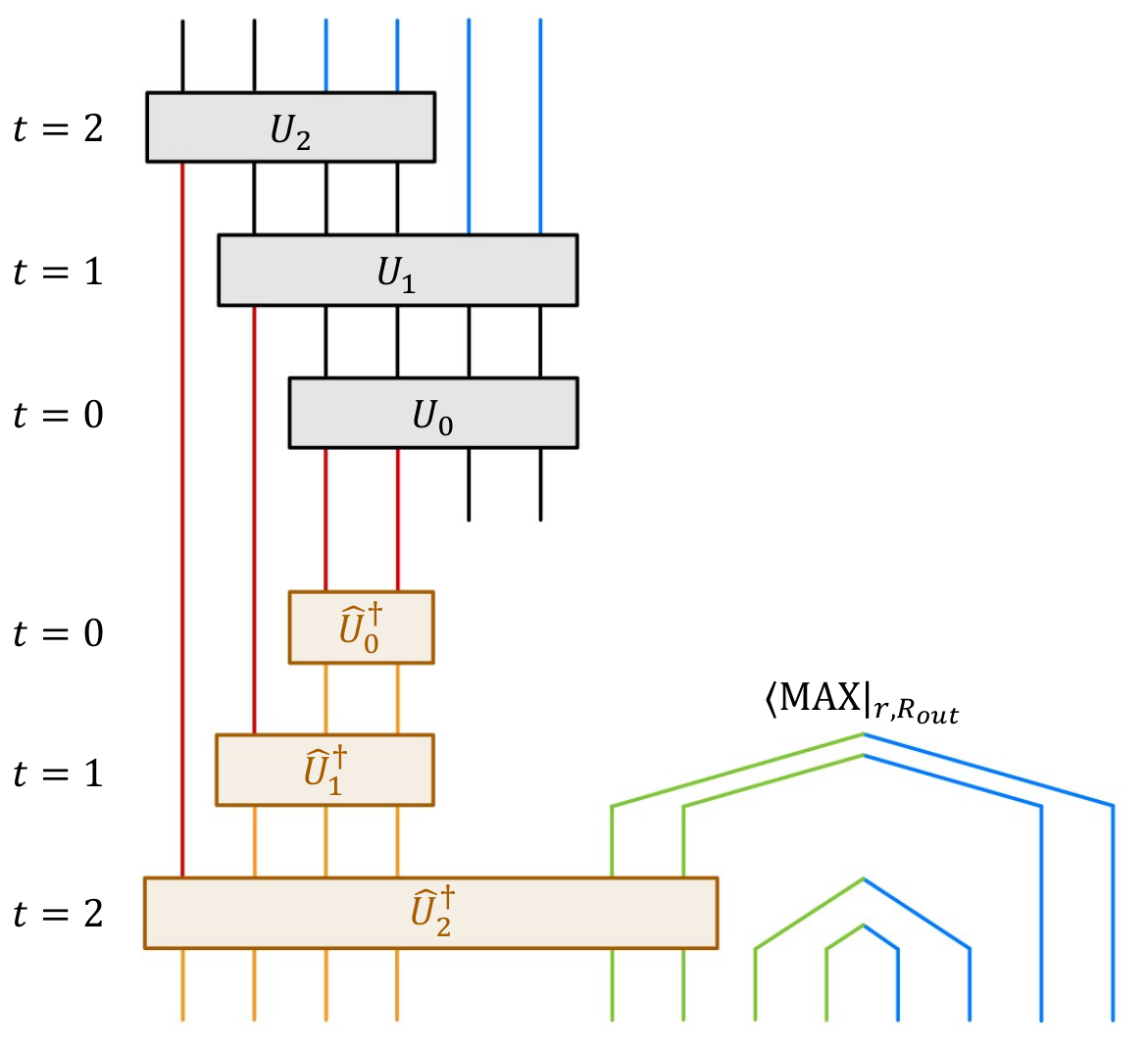}
    \caption{The  backwards-forwards holographic map relating the effective description with nontrivial dynamics at $t=2$ in figure~\ref{fig:eff_dyn_NEW} to the fundamental description at the same time in figure~\ref{fig:fun_dyn_PHEVA}, including backwards time evolution in the effective description, removal of the maximally entangled Hawking modes with $\langle \text{MAX}|_{r, R_\text{out}}$, and forward time evolution in the fundamental description.}
    \label{fig:V_NEW}
\end{figure}

Since any mode in $R$ belongs to both descriptions, we may let these $R_\text{in}$ degrees of freedom ``fall into'' the black hole in the fundamental description. This allows us to freely perform forwards time evolution in the fundamental description (after tensoring in the fixed states $f$) until time $t$. The output will be the degrees of freedom in $BR_\text{out}$, completing the holographic map:
\begin{equation}
\label{eq:BackwardsForwardsMap}
    V_t = U|\psi\rangle_f   \langle \text{MAX}|_{r, R_\text{out}} \hat{U}^\dagger\,,
\end{equation}
with $U \equiv U_t U_{t-1} \cdots U_1 U_0$ and $\hat{U} \equiv \hat{U}_t \hat{U}_{t-1} \cdots \hat{U}_1 \hat{U}_0$. We have the correct number of degrees of freedom automatically. Because the map involves backwards evolution in the effective description followed by forwards evolution in the fundamental description, we refer to it as the ``backwards-forwards'' holographic map. Figure~\ref{fig:V_NEW} shows an example of this holographic map for $m_0 = 2$, $n_0 = 4$, and $t=2$.

Let us consider the question of equivariance. Since the effective dynamics are no longer trivial, equivariance is now the statement that acting with the effective dynamics and then the holographic map is the same as acting with the holographic map and then the fundamental dynamics:
\begin{equation}
\label{eq:InteractingEquivariance}
    V_{t+1} \hat{U}_{t+1} = U_{t+1} V_t \,.
\end{equation}
It should be clear from the definition of the backwards-forwards map that it is equivariant by construction, since the first half of the map involves undoing effective dynamics and the second half performing fundamental dynamics; additional time evolution on the effective side is immediately undone by the map, with the corresponding fundamental time evolution added on at the end.  In equations, it is easy to verify that the definition (\ref{eq:BackwardsForwardsMap}) implies the equivariance condition (\ref{eq:InteractingEquivariance}). The automatic nature of equivariance is an attractive feature of this model.

\subsection{The backwards-forwards map as a post-selection map}

Consistency with the results of PHEVA requires that the backwards-forwards holographic map of figure~\ref{fig:V_NEW} should reduce to that of figure~\ref{fig:V_PHEVA} under the assumption that the effective dynamics are trivial, $\hat{U}_t = \mathbb{1}$. This is not apparently true at first glance, as the backwards-forwards map does not involve post-selection on the output of the fundamental dynamics. However, we can transform the backwards-forwards map to show that it is indeed equivalent to a post-selection map, 
when acting on states that dynamically evolved from matter falling into the black hole.
In the case of trivial effective dynamics, this post-selection map is the same as that of \cite{akers_black_2022}.

Although the backwards-forwards map does not have post-selection on the output of the fundamental dynamics, the annihilation of the Hawking pairs in the effective description with $\langle \text{MAX}|_{r,R_\text{out}}$ can be regarded as a kind of post-selection, with unit probability since the $rR_{\text{out}}$ modes are guaranteed to be in the maximally entangled state. The key observation is that we can transform one kind of post-selection into the other using the teleportation protocol, as in eq.~(\ref{eq:Teleportation}).

\begin{figure}
    \centering
    \includegraphics[width=\linewidth]{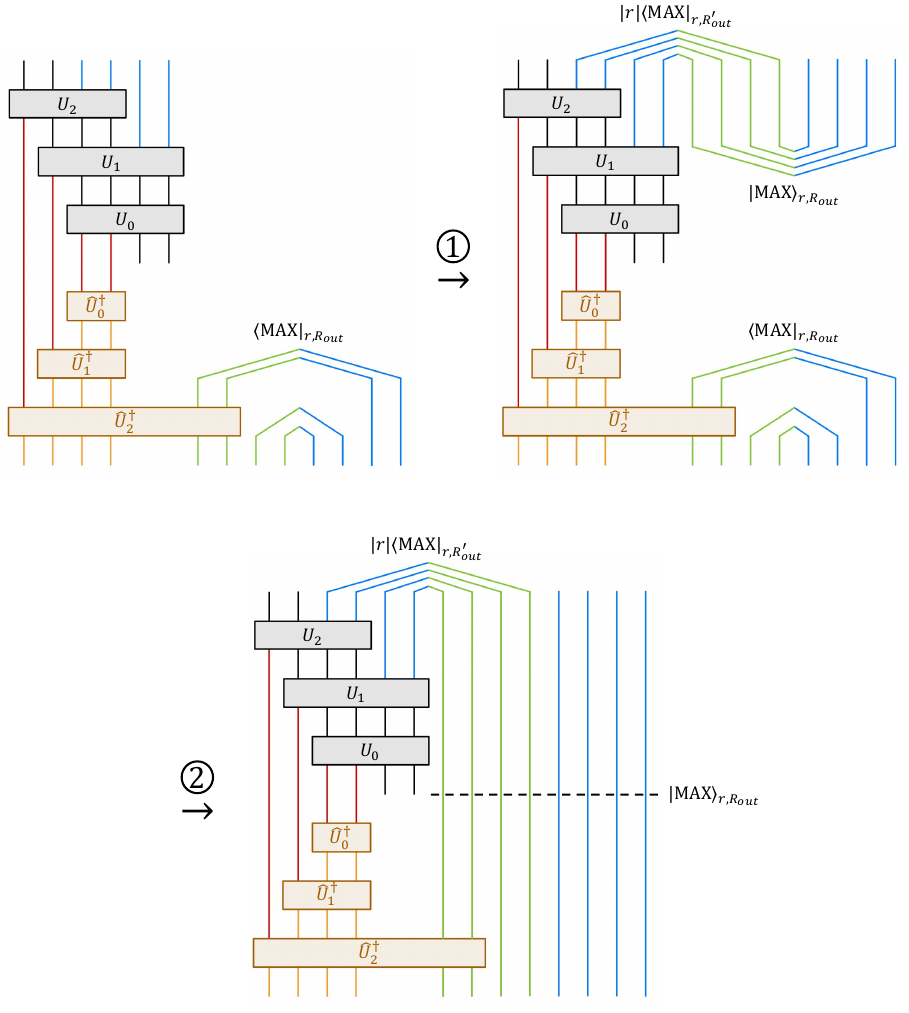}
    \caption{An illustration of the steps showing equivalence between the backwards-forwards holographic map and a post-selection map. The Hawking modes coming from the fundamental description are bent around into a teleportation protocol in step (1), and matched to the time-reversed entangled Hawking pairs from the effective description in step (2). At the dashed line, each qudit in $R_\text{out}$ is in the maximally entangled state $|\text{MAX}\rangle_{r,R_\text{out}}$ with its partner qudit in $r$; due to the interactions, this is not true in general at the bottom of the diagram.}
    \label{fig:BFtoPHEVA}
\end{figure}

A visual demonstration of these transformations is shown for $m_0=2$, $n_0=4$, and $t=2$ in figure~\ref{fig:BFtoPHEVA}. In step (1), we bend the lines of $R_\text{out}$ (output from the fundamental dynamics) into a teleportation protocol. There are now two copies of the $R_\text{out}$ Hilbert space in the later part of the map; as before, call the one coming out of the fundamental dynamics $R'_\text{out}$. This introduces the post-selection (with probability less than 1) on $R'_\text{out}$.

Next we observe that the annihilation of the Hawking pairs in the maximally entangled state of $rR_\text{out}$, followed by the reintroduction of modes in the same state, can simply be replaced by allowing the Hawking pairs to continue, becoming the entangled pairs in the teleportation protocol, before we post-select on the $r$ modes being maximally entangled with the radiation $R'_\text{out}$ coming from the fundamental dynamics.  Put another way, the entanglement insertion $|\text{MAX}\rangle_{r,R_\text{out}}$ required for the teleportation protocol combines with the annihilation of the Hawking modes $\langle \text{MAX}|_{r,R_\text{out}}$ in the effective description to form a projector onto the maximally entangled state, $|\text{MAX}\rangle\langle \text{MAX}|_{r,R_\text{out}}$, but because we know with certainty that the Hawking pairs emerge from the backwards effective dynamics in this entangled state, the projector acts as the identity. Therefore we replace the projector with connected lines in step (2).

After step (2), we see that the backwards-forwards holographic map has been re-cast to involve inserting fixed degrees of freedom $f$, acting with a unitary, and performing post-selection on the output of the fundamental dynamics. This is exactly the post-selection form proposed by PHEVA, with the unitary $U$ acting on $\ell rf$ given by both backwards and forwards time evolution,
\begin{equation}
    U = U_t U_{t-1}\dots U_0 \hat{U}_0^\dagger \dots \hat{U}_{t-1}^\dagger \hat{U}_{t}^\dagger.
\end{equation}
We note that in general, the input modes to this map in $rR_\text{out}$ are not in the maximally entangled state  $|\text{MAX}\rangle_{r,R_\text{out}}$ as they were in the non-interacting case of \cite{akers_black_2022} shown in  figure~\ref{fig:V_PHEVA}.
As described, at a general time $t$ the new interactions in the effective description will have spread the entanglement initially shared by $rR_\text{out}$ among the tripartite system $\ell rR_\text{out}$. It isn't until after we undo the effective dynamics that $R_\text{out}$ returns to a maximally entangled state with $r$. This is indicated in figure~\ref{fig:BFtoPHEVA} by the dashed line, at which point the $rR_\text{out}$ modes are in the maximally entangled state. It is this entanglement on $rR_\text{out}$ that combines with the post-selection on $rR'_\text{out}$ to form the teleportation protocol that teleports the information inside the black hole to the radiation.

\begin{figure}
    \centering
    \includegraphics[width=\linewidth]{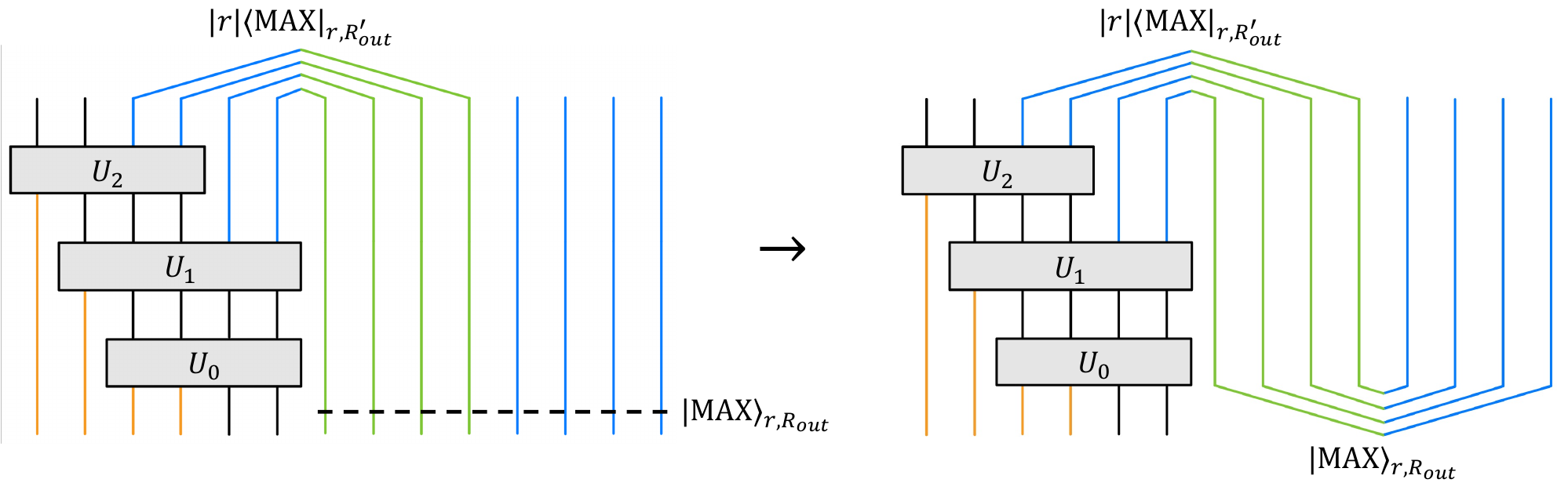}
    \caption{An illustration of obtaining the holographic map of \cite{akers_black_2022} from the backwards-forwards map in the limit of trivial effective dynamics. From the final map in figure~\ref{fig:BFtoPHEVA}, we set $\hat{U}_t=\mathbb{1}$, giving the left figure. Once the effective dynamics have disappeared, we can replace the dashed line indicating a maximally entangled state with the notation for entanglement insertion. The result matches figure~\ref{fig:V_PHEVA} exactly.}
    \label{fig:trivial_eff}
\end{figure}

In the case of trivial effective dynamics, we can show that this post-selection version of the backwards-forwards map reduces to precisely the PHEVA holographic map. This is illustrated in figure~\ref{fig:trivial_eff}. Replacing the $\hat{U}_t$ with identities, the effective dynamics disappear from the holographic map. This allows us to reintroduce the entanglement insertion $|\text{MAX}\rangle_{r,R_\text{out}}$ found at the bottom of figure~\ref{fig:V_PHEVA} since no effective dynamics have changed the entanglement of $rR_\text{out}$. The map then reduces exactly to figure~\ref{fig:V_PHEVA} -- the backwards-forwards holographic map proposed here is indeed compatible with the post-selection map proposed by \cite{akers_black_2022}. 
For the case with additional interactions, our model represents a suitable generalization.

The steps performed above to relate the backwards-forwards map to the post-selection map were only possible because we knew that the inputs came from a valid black hole state in the effective description
-- one that formed dynamically from non-singular matter falling into the black hole. 
Had this not been the case and we applied the backwards-forwards map to any generic state in the effective description, the post-selection $\langle \text{MAX}|_{r,R_\text{out}}$ in figure~\ref{fig:V_NEW} would not succeed with probability 1, 
and replacing the projector $|\text{MAX}\rangle \langle \text{MAX}|_{r,R_\text{out}}$ with the identity in step (2) would not be possible. In other words, the backwards-forwards map by itself is not unitary -- it will not preserve all inner products when acting on generic states. However, the backwards-forwards holographic map appears to be unitary when acting only on valid black hole states in the sense that it preserves the inner products of these states.

\section{Interactions outside the black hole and complexity}
\label{sec:outside}

The effective and fundamental descriptions share the same description of the physics outside the black hole, and the holographic map should act trivially on the reservoir $R$. In this section we consider adding interactions between modes outside the black hole. We show that the backwards-forwards map continues to act as the identity outside the horizon when such interactions take place, and in fact that the map is independent of the interactions.

\subsection{Interactions outside the black hole} 

\begin{figure}
    \centering
    \includegraphics[width=0.9\linewidth]{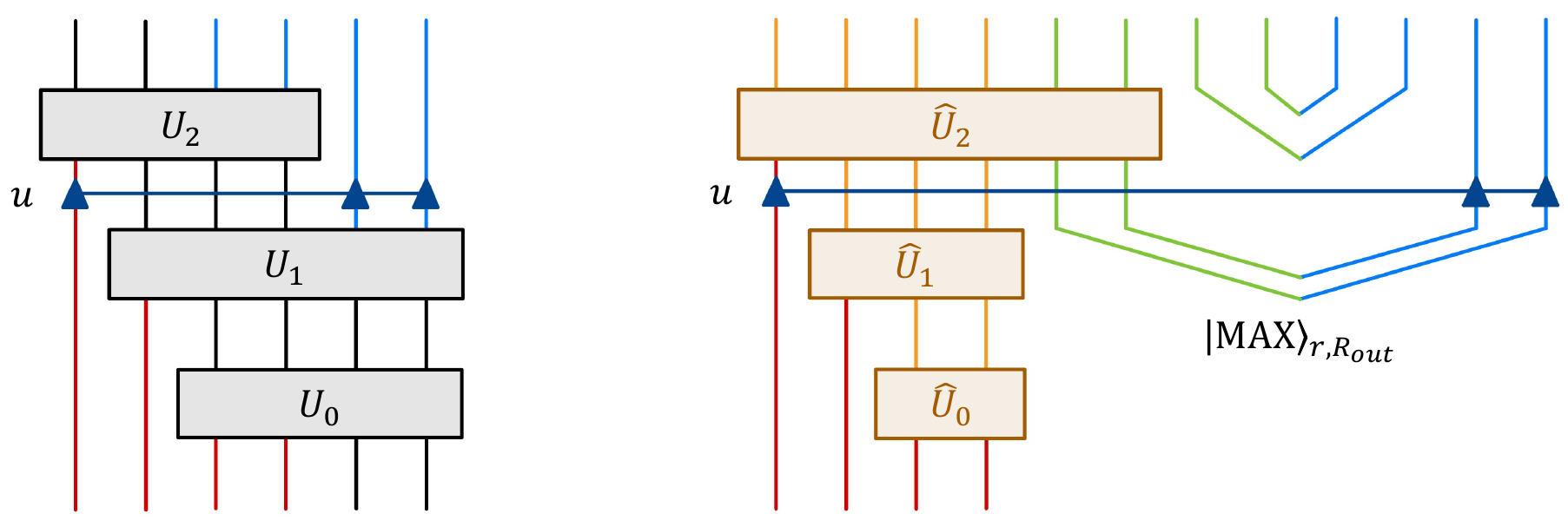}
    \caption{Fundamental (left) and effective (right) dynamics including interactions between $R_\text{in}$ and $R_\text{out}$ for $m_0=2$ and $n_0=4$ up until $t=2$. Connected dark blue triangles denote a unitary operator $u$ acting on those qudits.}
    \label{fig:dyn_Rout}
\end{figure}

Let us consider interactions between modes outside the black hole. For example, as a qudit in $R_\text{in}$ falls towards the black hole, it may interact with already-emitted Hawking radiation in $R_\text{out}$ before crossing the horizon. These interactions are the same in both the effective and fundamental descriptions since they occur in the reservoir $R$. In circuit diagrams we will denote these unitaries $u$ by placing triangles on the interacting qudits, following the notation of \cite{kim_complementarity_2022}. Figure~\ref{fig:dyn_Rout} shows such an interaction in the fundamental and effective dynamics for $m_0=2$ and $n_0=4$ up until $t=2$.

\begin{figure}
    \centering
    \includegraphics[width=0.6\linewidth]{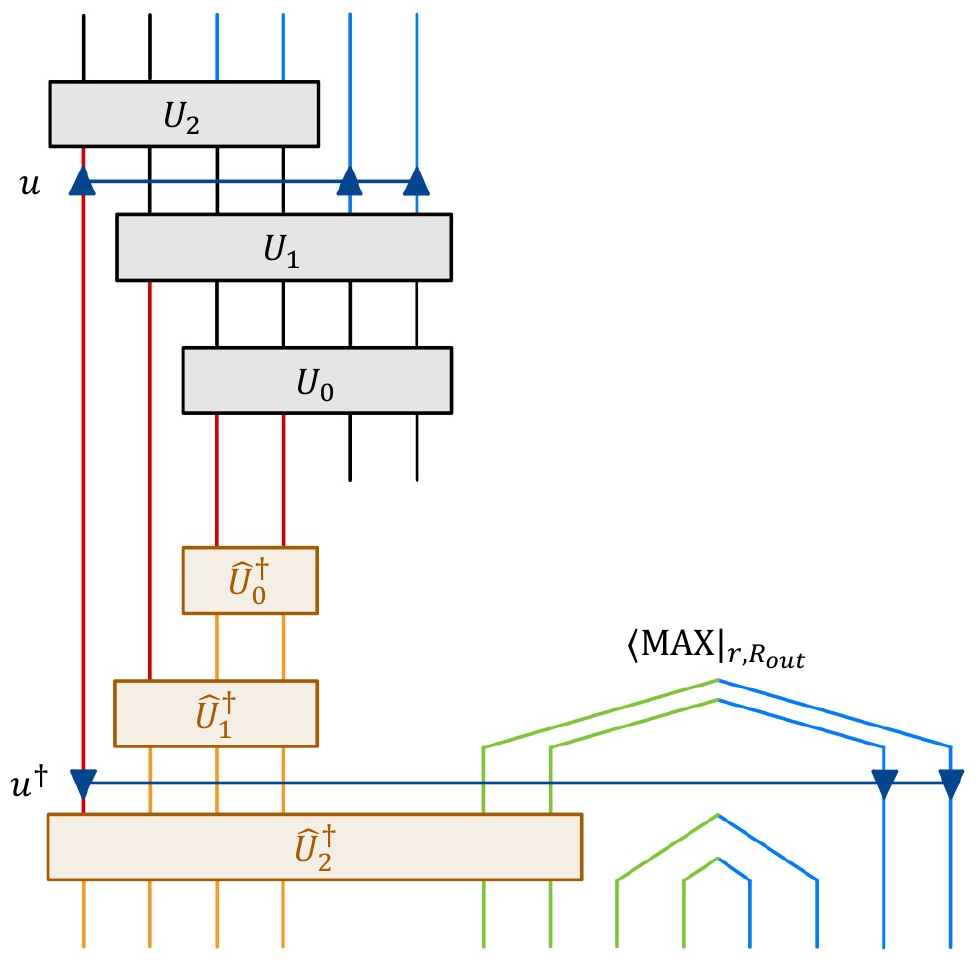}
    \caption{The backwards-forwards holographic map including interactions between infalling qudits in $R_\text{in}$ and outgoing radiation in $R_\text{out}$.}
    \label{fig:V_Rout}
\end{figure}

As described in section~\ref{sec:review}, the holographic map on black hole interiors should act trivially on the radiation $R_\text{out}$ outside the black hole. Let us see whether our construction of the backwards-forwards map satisfies this expectation. If we keep the principle that the backwards-forwards map includes all interactions, then as we time evolve backwards in the effective description, we also undo the $R_\text{out}$ interaction using $u^\dagger$; similarly, we include the $R_\text{out}$ interaction $u$ in the subsequent forwards fundamental evolution. Figure~\ref{fig:V_Rout} shows the updated backwards-forwards holographic map with the $R_\text{out}$ interactions.

\begin{figure}
    \centering
    \includegraphics[width=\linewidth]{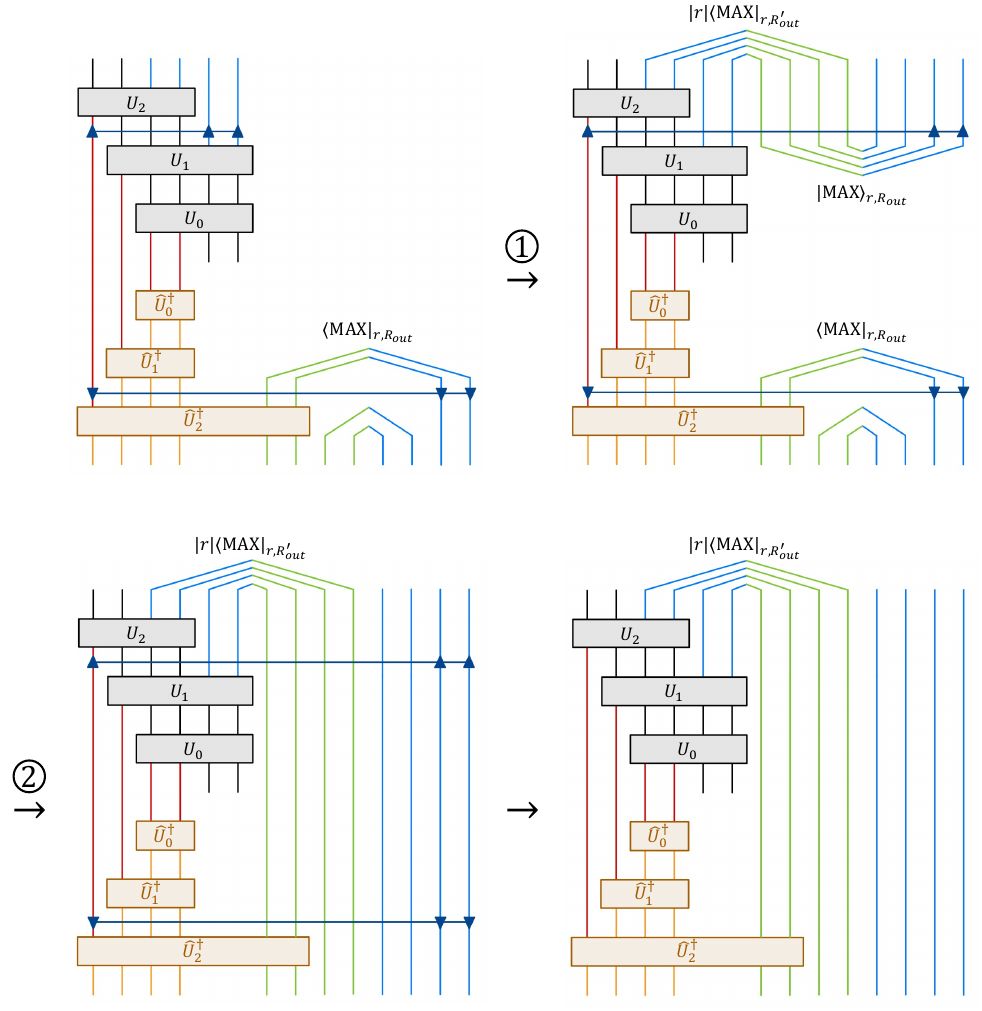}
    \caption{Transformations between the backwards-forwards holographic map and its post-selection form, showing that interactions between $R_\text{in}$ and $R_\text{out}$ outside the black hole vanish from the holographic map.}
    \label{fig:BFtoPHEVA_Rout}
\end{figure}

It is not immediately obvious from figure~\ref{fig:V_Rout} that including $R_\text{out}$ interactions in the backwards-forwards map satisfies the principle of trivial action on the reservoir, since  the $u^\dagger$ and $u$ interactions act on disconnected lines in $R_\text{out}$. However, we can resolve the issue by transforming the holographic map into the equivalent post-selection form, as described in section~\ref{sec:BFmap} in figure~\ref{fig:V_Rout}.
These transformations are shown with the $R_\text{out}$ interactions in figure~\ref{fig:BFtoPHEVA_Rout}. After step (1), the portion of the unitary $u$ acting on the radiation in the fundamental description may be moved around the teleportation protocol. After connecting the $rR_\text{out}$ lines in step (2), $u$ and $u^\dagger$ 
act on the same lines. They commute with the intervening $U_t$ and $\hat{U}_t$ unitaries associated with earlier time steps, since those unitaries act on modes already at or beyond the horizon. Thus we may combine $u$ and $u^\dagger$ to the identity  -- the $R_\text{out}$ interactions disappear completely from the holographic map.

Thus the backwards-forwards map indeed acts trivially on the radiation, satisfying our expectation. In fact, we have shown that the holographic map does not depend on the interactions outside the black hole at all -- at first we included them as part of the backwards-forwards dynamics, but since they are the same in both descriptions they drop out, and we end up with the same holographic map as if there had been no interactions outside the black hole at all.

\begin{figure}
    \centering
    \includegraphics[width=\linewidth]{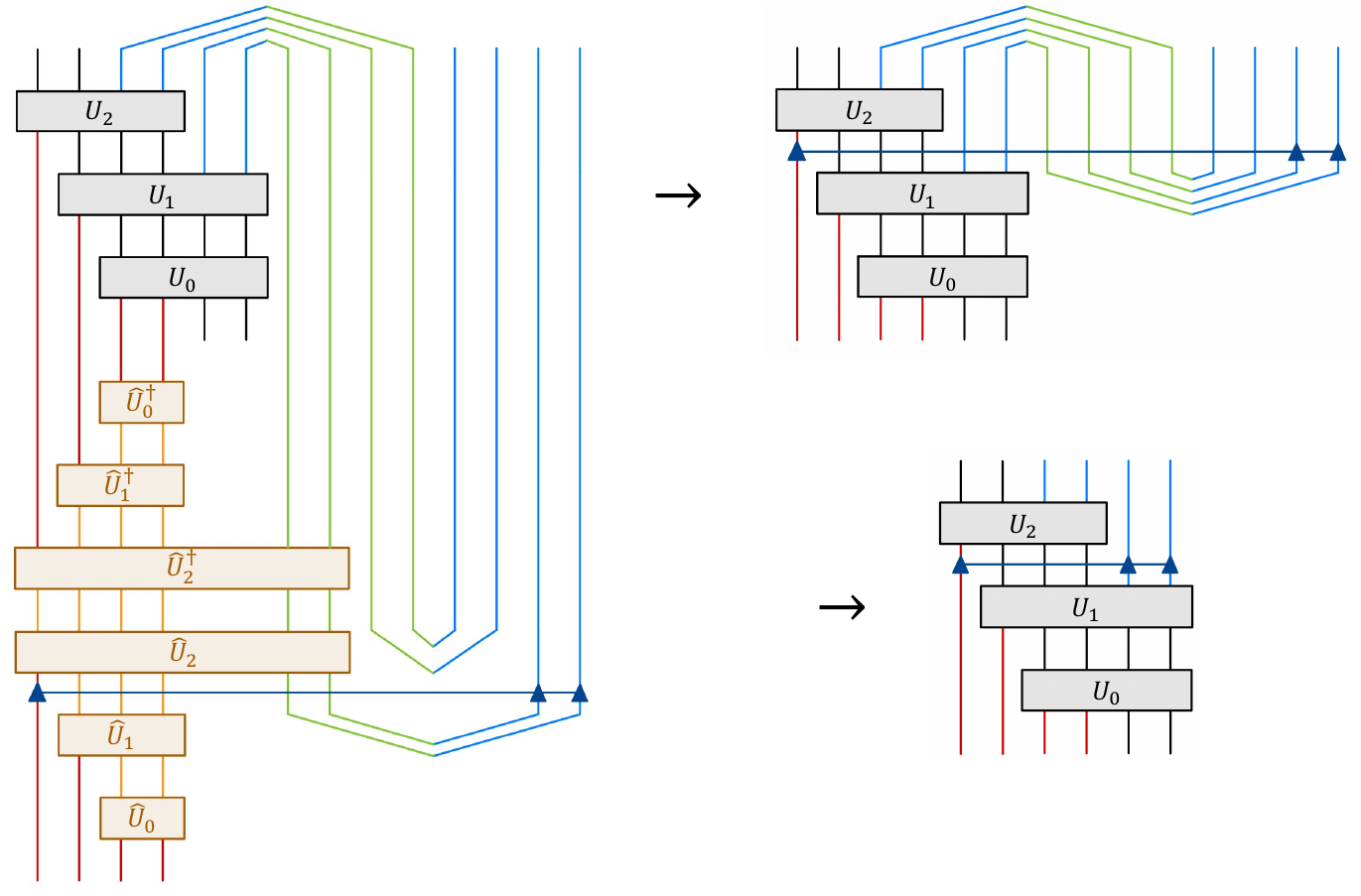}
    \caption{Demonstrating equivariance is satisfied by the post-selection version of the new holographic map in the presence of interactions outside the horizon. The left figure shows effective time evolution composed with the holographic map found in figure~\ref{fig:BFtoPHEVA_Rout}. The $|\text{MAX}\rangle$ notation has been omitted for ease of reading.}
    \label{fig:equi}
\end{figure}

We can also verify that equivariance continues to hold  when we include the $R_\text{in}R_\text{out}$ interactions, as shown in figure~\ref{fig:equi}. If we compose the effective dynamics of figure~\ref{fig:dyn_Rout} with the post-selection holographic map found in figure~\ref{fig:BFtoPHEVA_Rout}, all unitary dynamics $\hat{U}_t$ in the effective description at time steps later than the reservoir interaction contract with their inverses $\hat{U}^\dagger_t$ and reduce to the identity. Since the reservoir interaction $u$ commutes with the effective $\hat{U}_t$ and fundamental $U_t$ unitaries at earlier times, nothing obstructs it from moving into the upper part of the diagram to the same time step amongst the fundamental interactions. The remaining $\hat{U}_t$ and $\hat{U}_t^\dagger$ then cancel, leaving us with the second diagram in the figure. Straightening the $rR_\text{out}$ lines we find fundamental dynamics with the correct $R_\text{in}R_\text{out}$ interactions, matching those shown in figure~\ref{fig:dyn_Rout}, and the map is equivariant. 

\subsection{Unitarity and computational complexity} \label{sec:unitarity}

We turn now to the questions posed by Kim and Preskill in \cite{kim_complementarity_2022} concerning these non-isometric holographic maps. First, they asked whether interactions between infallers and $rR_\text{out}$ modes in the effective description could combine with the post-selection in these maps to cause violations to the unitarity of black hole evaporation. To understand this, the work of \cite{kim_complementarity_2022} capitalizes on a particular property of PHEVA's holographic map: straightening the lines in figure~\ref{fig:V_PHEVA} reveals the fundamental dynamics. After adding an infaller with interactions, straightening those lines leads to partially transposed unitaries that could pose a threat to unitarity. Averages over the Haar measure in the fundamental dynamics showed that these partial transpositions only lead to exponentially suppressed deviations from unitarity. This indicated that only small corrections to the PHEVA holographic map should be needed to restore unitary black hole evolution.

By including all effective dynamics among $\ell rR$ in our backwards-forwards holographic map, we restore complete unitarity to black hole evaporation. Straightening the lines of figure~\ref{fig:V_PHEVA} can be viewed as equivalent to running the transformations of figure~\ref{fig:BFtoPHEVA} backwards in our model, converting the post-selection type map back to the backwards-forwards holographic map. This process does not lead to any partially transposed unitaries that could threaten the unitarity of the fundamental description; in our construction the external interactions $u$ are included along with their inverses and cancel, or equivalently need not be included at all. 
In addition, as noted at the end of section~\ref{sec:build}, the backwards-forwards map appears to be unitary when we act on valid black hole states in the effective description. It is only when we act on more general states (for which the holographic map is not built for) that we see violations of unitarity from the post-selection representing the annihilation of Hawking pairs.

Second, post-selection is known to lead to other unexpected or undesirable effects in quantum circuits. For example, it has been shown that the class of problems solvable by adding post-selection to a polynomial depth quantum circuit is a very large class, containing NP \cite{aaronson_quantum_2005}. Motivated by the black hole final state proposal of \cite{horowitz_black_2004}, \cite{bao_grover_2016} demonstrated that post-selection can lead to superluminal signaling and speedups in search exceeding Grover's algorithm. Kim and Preskill then asked whether the post-selection in the holographic maps described by PHEVA could lead to superpolynomial computational complexity. They were able to show that the complexity of the map was instead limited by the complexity of the infaller, which can remain subexponential.

Our backwards-forwards map helps to resolve these issues for holographic encodings of black hole interiors. Importantly, the post-selection present in the backwards-forwards map is special: for valid black hole states on which it is designed to act, it is successful with probability 1. There is effectively no post-selection since it is guaranteed to succeed. Thus it cannot lead to an exponential increase in computational complexity, superluminal signaling, speedups in Grover's algorithm, or other issues related to post-selection. Furthermore, the post-selection holographic maps of \cite{akers_black_2022} are related to the backwards-forwards map by the transformations of figure~\ref{fig:BFtoPHEVA}. This may be taken as a demonstration that those post-selection maps are free of these concerns as well. Constructing the holographic map using our backwards-forwards prescription shows that encodings of black hole interiors are not plagued by problems typically associated with post-selection.

\section{Conclusion} \label{sec:conc}

After introducing non-trivial dynamics among effective degrees of freedom, we have demonstrated a new construction for non-isometric holographic maps encoding black hole interiors. This ``backwards-forwards'' map (shown in figure~\ref{fig:V_NEW}) takes dynamics in both descriptions into account by performing backwards  time evolution in the effective description of the black hole followed by forwards time evolution in the fundamental description. Furthermore, we have given a series of transformations (depicted in figure~\ref{fig:BFtoPHEVA}) re-expressing this map as a post-selection holographic map of the sort proposed in \cite{akers_black_2022}, and reducing to their construction in the case of trivial effective dynamics.  Our holographic map is equivariant with respect to time evolution, and is independent of any interactions outside the black hole.

Furthermore, we considered potential violations to the unitarity of black hole evaporation and the possibility of exponential computational complexity in our new construction, following the work of \cite{kim_complementarity_2022}. Thanks to the inclusion of effective dynamics in our new holographic map, an interacting infaller no longer poses any threat to the unitarity of the black hole $S$-matrix. In addition, because the backwards-forwards map involves post-selection with probability 1, it cannot lead to superpolynomial computational complexity, superluminal signaling, or speedups to Grover's algorithm. These results are exact and do not require averaging or infallers with restricted complexity.

Future work is needed to better understand the effective dynamics $\hat{U}_t$. For the purposes of this work, we have imagined that the unitary interactions described by $\hat{U}_t$ are as general as possible: $\hat{U}_t$ acts globally on all qudits inside the black hole. Because the effective description involves the geometry of spacetime, it could be possible to improve upon this by taking some form of locality into account. Some distant $\ell$ mode that crossed the horizon a long time ago may interact very weakly (if at all) with a new $R_\text{in}$ degree of freedom just crossing the horizon. Additionally, as this is only an \textit{effective} description of the black hole, it may be possible for $\hat{U}_t$ to only be approximately unitary with exponentially suppressed deviations from unitarity. The results described in this paper do not require any special properties of $\hat{U}_t$ beyond unitarity, so other resources are needed to better understand its nature. Matching the backwards-forwards holographic map to a well understood AdS/CFT dictionary might offer some clues in this direction. What's more, including effective dynamics in the holographic map shouldn't invalidate PHEVA's complexity results in \cite{akers_black_2022}; ensuring their results hold here might offer further constraints on the effective dynamics.

Throughout this work we have restricted to the subset of the effective description Hilbert space that is accessible by unitary effective dynamics from an initial state of matter falling into the black hole, that is, the states that are in one-to-one correspondence with states in the fundamental description. This subset of states is characterized by the initiation of Hawking pairs in the maximally entangled state $|\text{MAX}\rangle_{r,R_\text{out}}$, and the restriction to these states was necessary to demonstrate the equivalence of the backwards-forwards and post-selection maps, in replacing a projector to maximally entangled states with the identity map.  However, there are situations where it is natural to consider all generic states of the effective Hilbert space; for example, a measurement theory for the infalling observer would naturally include projectors onto all states, not just those that are dynamically accessible. A generic state of the effective description, evolved backward in time, would instead reach a past singularity; these states will not have the correct entanglement between $rR_\text{out}$ and will naively be annihilated by the backwards-forwards map, but their singular nature makes this conclusion delicate. It was shown that both the generic and dynamical maps of \cite{akers_black_2022} were good for all subexponential states, including those with past singularities, and thus a better understanding of how to characterize the action of the backwards-forwards map on such states is an important step for future work.

\acknowledgments

We are grateful to Sristy Agrawal, Chris Akers, Daniel Harlow, Joshua Levin,  and Graeme Smith for helpful discussions. The authors are supported by the Department of Energy under grants DE-SC0010005 and DE-SC0020360.



\bibliographystyle{JHEP}
\bibliography{biblio.bib}






\end{document}